\documentclass[naturemag,twocolumn]{revtex4-1}
\usepackage{graphicx}
\usepackage{dcolumn}
\usepackage{bm}
\usepackage{color}
\usepackage{soul}
\usepackage[english]{babel}

\begin{document}

\newcommand{\soleil}{$^2$}
\newcommand{\phlam}{$^1$}
\title{Coherent Terahertz synchrotron radiation mastered by controlling the irregular dynamics of relativistic electron bunches}
\author{C. Evain$^*$\phlam, C. Szwaj\phlam, E. Roussel\phlam, J. Rodriguez\phlam, M. Le Parquier\phlam,  M.-A. Tordeux\soleil, F. Ribeiro\soleil, M. Labat\soleil, N. Hubert\soleil, J.-B. Brubach\soleil,P. Roy\soleil, S. Bielawski\phlam}
\affiliation{\phlam Univ. Lille, CNRS, UMR 8523 - PhLAM - Physique des Lasers, Atomes et Mol\'ecules,  Centre    d'\'Etude Recherches et Applications (CERLA), F-59000 Lille, France.\\
  \soleil Synchrotron SOLEIL, Saint Aubin, BP 34, 91192 Gif-sur-Yvette, France}
\email[Corresponding author : ]{clement.evain@univ-lille.fr}
\date{\today}
\maketitle

{\bf Spontaneous formation of spatial structures (patterns) occurs in various contexts, ranging from sand dunes~\cite{charru2013sand} and rogue
  wave formation~\cite{hopkin2004sea,solli2007optical}, to traffic jams~\cite{helbing2001traffic}.
  These last decades, very practical reasons also led to studies of pattern formation in relativistic electron bunches used in synchrotron radiation light sources.
  As the main motivation, the patterns  which spontaneously appear during an instability  increase the terahertz radiation power by  factors exceeding 10000~\cite{abo2002.PhysRevLett.88.254801,byrd2002.PhysRevLett.89.224801}.
  However the irregularity  of these patterns~\cite{byrd2002.PhysRevLett.89.224801,venturini2002.PhysRevLett.89.224802, abo2002.PhysRevLett.88.254801,uvsor_MBI_ybco,roussel2015.EOS,brosi2016fast,PhysRevAccelBeams.19.020704} largely prevented applications of this powerful source.
  Here we show how to make the spatio-temporal patterns regular (and thus the emitted THz power) using a point of view borrowed from chaos control theory~\cite{ott1990controlling,shinbrot1993using,pyragas1992continuous}.
   Mathematically, regular unstable solutions are expected to coexist with the undesired irregular solutions, and may thus be controllable using
   feedback control. We demonstrate the stabilization of such regular solutions in the Synchrotron SOLEIL storage ring. Operation of these “controlled
   unstable solutions” enables new designs of high charge and stable synchrotron radiation sources.}

\begin{figure*}[htbp!]\center
  \includegraphics[width=0.94\linewidth]{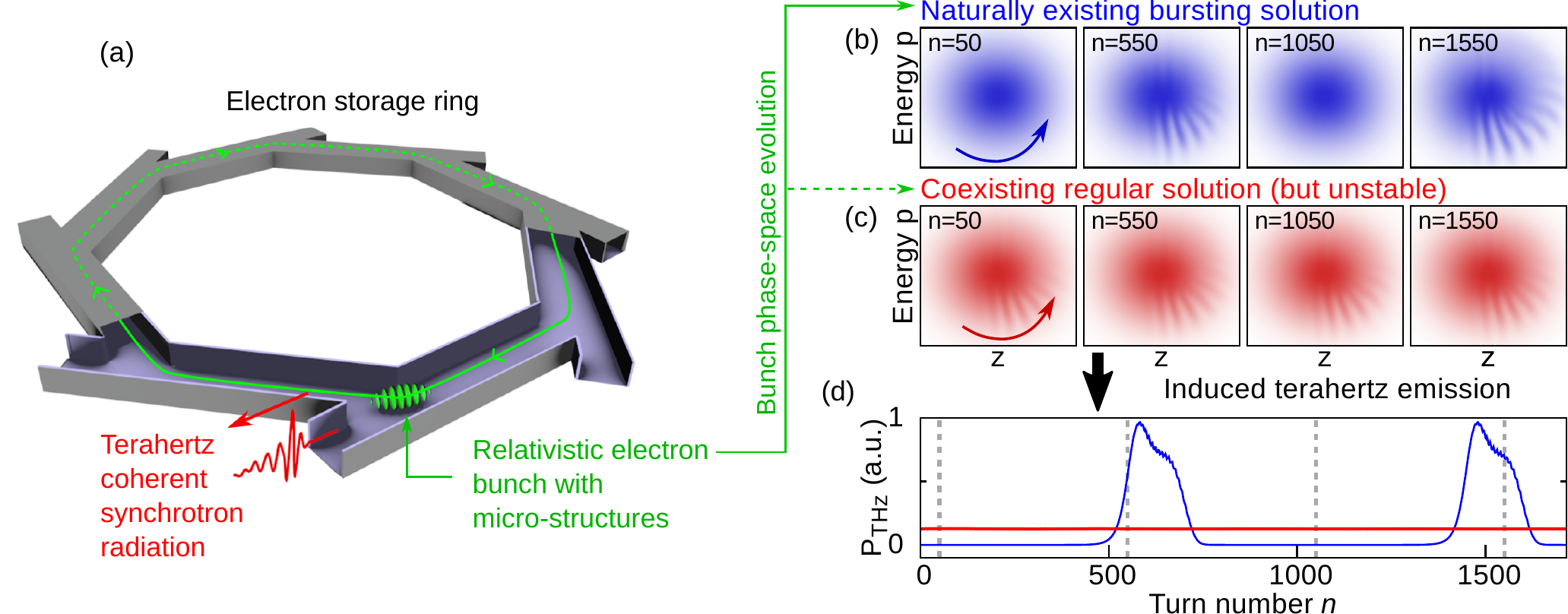}
  \caption{{\bf Storage ring synchrotron facilities, and the microbunching instability}.(a): Typical layout of a synchrotron radiation facility as SOLEIL. One (or several) relativistic electron bunches are stored on a closed-loop  trajectory. Above a threshold charge density a microstructure  spontaneously appears inside the bunch as the result of a spatio-temporal instability arising from the interaction between  electrons. The microstructure emits a huge power of coherent light,  typically in the THz range. (b,c,d): Numerical results in the case   of SOLEIL. (b) electron bunch shapes evolution in phase-space at   four different times of the bursting dynamics. (c): electron bunch  shape corresponding to an unstable and regular solution existing in   phase space, but which can not be observed experimentally (because   of its unstable nature). Our feedback control strategy will aim at   stabilizing this solution. (d) terahertz emission induced by the   structures, as would be recorded by a bolometer. Blue: natural   bursting solution. Red: terahertz emission corresponding to the   regular and unstable solution (c). See Methods for simulation and parameters details.}
  \label{fig:1}
\end{figure*}

Synchrotron light sources are used worldwide to produce brilliant light from  THz to hard X-rays, allowing to investigate a very large range of matter properties.
In these sources where electron bunches travel at relativistic velocities, an ubiquitous phenomenon occurs when the bunch charge density exceeds a threshold value.
Due to the interaction between the electron bunch and its own emitted electric field, micro-structures spontaneously appear in the longitudinal profile (and phase-space) of the bunch~\cite{stupakov2002.PhysRevSTAB.5.054402,venturini2002.PhysRevLett.89.224802,sannibale2004.PhysRevLett.93.094801,Warnock:2006qa} (see Fig.~\ref{fig:1}(a) and (b) for the SOLEIL storage ring which will be considered here).
In storage rings, these structures are responsible for a huge emission of coherent light in the terahertz range, typically $10^3-10^5$ times normal synchrotron radiation power density.
However, as the micro-structures appear mostly in the form of bursts [Fig.~\ref{fig:1}(d)]
this type of source is barely usable in user applications. Hence, research in this domain has naturally attempted to find regions for
which coherent emission (CSR) occurs while bursting dynamics is absent. Such
"parameter search" methods succeeded in identifying parameter regions with stable coherent emission. However, this corresponds to
special configurations (with short and low charge electron bunches, in the so-called low-alpha operation~\cite{abo2002.PhysRevLett.88.254801,byrd2002.PhysRevLett.89.224801,shimada2009transverse,feikes2011metrology,martin2011experience,roy.RSI.84.033102.2013,PhysRevAccelBeams.19.020704,brosi2016fast,tammaro2015high,steinmann2016frequency}) which are not compatible with most of the user experiments.
Therefore, this type of THz source is used in relatively few synchrotron radiation facilities (SOLEIL, DIAMOND, BESSY-II), and only during a small part of the year.
 
"Parameter search approaches" are however not the only possibilities for avoiding instabilities. The point of view that we will use here is directly borrowed from the so-called chaos control theory, introduced by Ott, Grebogi and Yorke (OGY)~\cite{ott1990controlling,shinbrot1993using}. Mathematically, when an undesired irregular solution is observed -- bursts of microbunching in our specific case -- other unstable solutions usually coexist. The existence of these unstable solutions is often viewed  as mathematical curiosities. However, OGY pointed out  two properties
with far-reaching applications. First {\it these co-existing unstable solutions are generically controllable}, i.e., there exists feedback loops between an observable and available parameters, that can force the system to stay onto the unstable solution. Furthermore, {\it as the feedback stabilizes an already existing solution, the extra-power  required for stabilization is theoretically infinitesimally small} (it is only limited by imperfections or noise, see Refs.~\cite{shinbrot1993using,boccaletti2000control} for details). These two points make the OGY strategy a good candidate for the control of high power systems,  like synchrotron radiation facilities.
  
\begin{figure*}[htbp!]\center
  \includegraphics[width=0.93\linewidth]{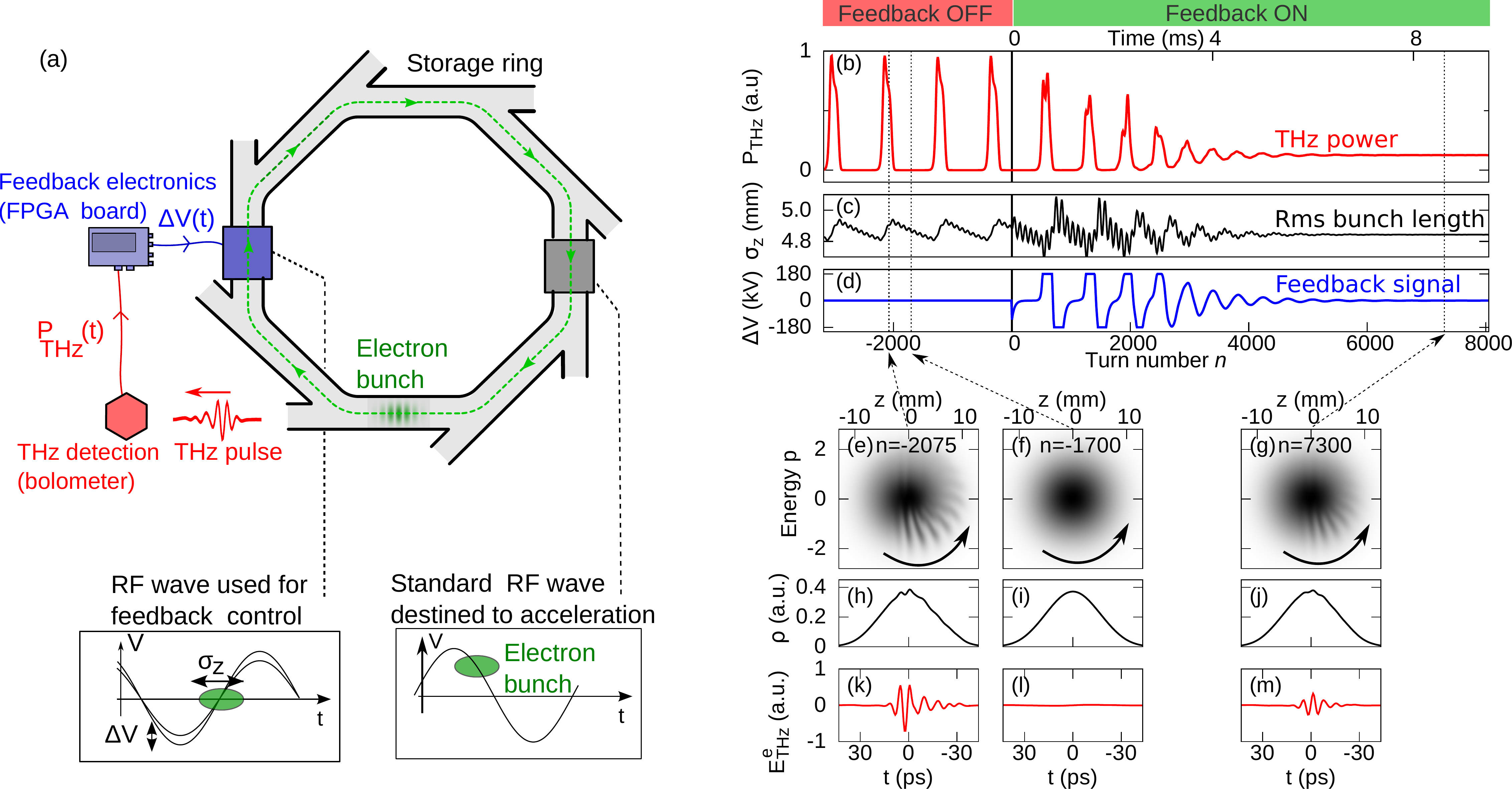}
  \caption{{\bf Control of the micro-bunching instability at SOLEIL: experimental setup and results expected numerically.} (a) experimental setup. The THz power $P_{\mathrm{THz}}(t)$ emitted by the
      electron bunch is monitored by a bolometer. A Field Programmable Gate Array (FPGA) computes the feedback control signal $\Delta V(t)$ in real time.
      This signal modulates the amplitude of the 352.2~MHz RF wave injected in one of the accelerating cavities. Right: numerical results obtained when the feedback control is applied at
      $t=0~$ms. (b-d), temporal evolution of the THz power $P_{\mathrm{THz}}(t)$, bunch length $\sigma_z(t)$, and feedback signal $\Delta V(t)$. (e-m): Electron bunch phase-space, bunch shape $\rho(z)$, and coherent THz pulse near the electron-bunch $E_{\mathrm{THz}}^e(z)$, at three times (see also the movie for the entire dynamics, and Methods for the numerical details). Lower left insets: relationship between the electron bunch and the power RF waves injected in the control accelerating cavity, and in normal accelerating  cavities.}      
  \label{fig:2}
\end{figure*}

In this Letter, we show that it is possible to suppress the bursts of microbunching instabilities occurring in storage-ring synchrotron
radiation facilities, using this strategy inspired from chaos control. We will demonstrate that, although the acceleration powers used in
synchrotron radiation facilities are in the fraction-of-MegaWatt range, tiny actions on a control parameter are sufficient to maintain the system
 on an unstable regular solution of phase-space, as the one displayed in Fig.~\ref{fig:1}c.

The experimental design of the feedback loop, and simulated expected performances are summarized in Figure~\ref{fig:2}. A fast bolometer
(with 1~$\mu$s response time) measures the power fluctuations $P_{\mathrm{\mathrm{THz}}}(t)$ of the emitted coherent THz radiation. This signal is used for
calculating a control signal $\Delta V(t)$, using a Field-Programmable Gate Array (FPGA). The control signal is then directly applied onto
one of the four accelerating cavities of the storage ring (see Figure~\ref{fig:2}a and Methods for details).
As detailed after, the action on the RF cavity signal is to modify the bunch length, which is directly linked to the bursting behaviour (see Fig.~2c for $t<0$).
In this first study, the feasibility of control is tested using only elementary feedback loops,  and in particular we have performed a systematic study using the so-called Pyragas method, which has proven to be efficient for stabilizing various non-accelerator systems~\cite{boccaletti2000control, Ahlborn2004}.
In numerical simulations and then with the FPGA used in the experiment, we compute the following signal $\Delta V(t)$:
\begin{eqnarray}
&& \frac{dX(t)}{dt}=\frac{1}{\tau_{LP}}\left[P_{\mathrm{THz}}(t)-X(t)\right]\label{eq:feedback2}\\
&&\Delta V(t)= G \left[X(t) - X(t-\tau) \right]\label{eq:feedback1},
\end{eqnarray}
where $P_{\mathrm{THz}}$ is the signal detected by the bolometer, at the THz and infrared AILES beamline~\cite{roy.IPT.49.139.2006}. Equation~(\ref{eq:feedback2}) represents a
low-pass filter with a time constant  $\tau_{LP}$. Equation~(\ref{eq:feedback1}) represents the Pyragas feedback, which involves two parameters $G$ (gain) and $\tau$ (delay). $\Delta V$
corresponds to the voltage amplitude applied to the accelerating cavity of the storage ring. The cutoff frequency of the low-pass
filter is 3~kHz, i.e., slightly lower that the bandwidth of the high power RF system (10~kHz). This filter removes components at the
revolution frequency (845 kHz), as well as the frequency due to the rotation of the microstructures in phase-space~\cite{evain2017direct}, in the 80~kHz range here. In order to find the conditions for stabilization, the feedback parameters $G$ and $\tau$ have been systematically scanned, while monitoring the remaining fluctuations of the THz power (both in the
numerical design study, and in the experiment). 

A typical numerical simulation result is displayed in Figures~\ref{fig:2}(b-m) (see Methods for numerical details).
Figure~\ref{fig:2}(b) represents a transient that occurs when the feedback is applied, showing that the bursts of terahertz power are
expected to be suppressed. However the microstructure pattern still exists in phase-space (see Fig.~2g), leading to a regular emission of coherent
terahertz radiation. Moreover, we can see that the feedback signal $\Delta V(t)$ tends to extremely small values after the transient has
died out (Fig.~2d). This is a consequence of the fact that we are stabilizing an {\it already existing} periodic state of the system.

Numerical simulations also showed that efficient control is expected when the control voltage amplitude $\Delta V(t)$ acts on the {\it slope} of the
accelerating field. Thus increasing $\Delta V(t)$ should only "compress" the bunch and vice-versa (see left insets of Fig.~2). However, a direct  variation of the RF signal amplitude
will also accelerate or decelerate the bunch (more precisely the so-called synchronous electron~\cite{sannibale2004.PhysRevLett.93.094801}). 
This is a key point in the realization of an efficient feedback control, and it led us to design the experiment so that the feedback signal $\Delta V(t)$ is applied in an
accelerating cavity operating in the so-called "zero-crossing" situation (see Methods).

\begin{figure}[htbp!]\center
  \includegraphics[width=1\linewidth]{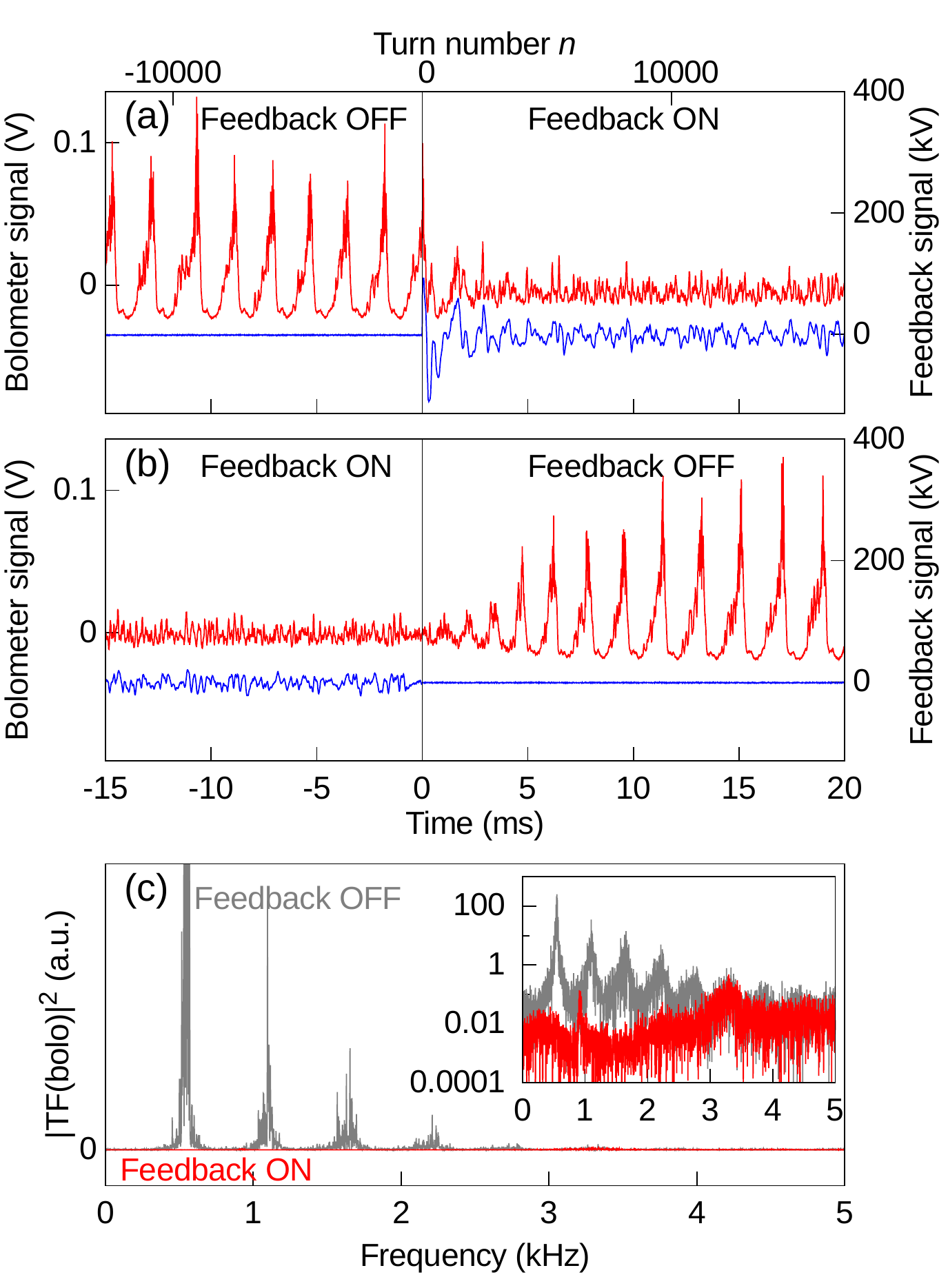}
  \caption{{\bf Feedback control of the micro-bunching instability:
    experimental results}. (a) and (b) transients observed when the
    feedback control is suddenly turned ON and OFF respectively. Red
    curves correspond to the bolometer voltage evolution, and blue
    curves correspond to the feedback control signal (amplitude
    modification of the signal in the control cavity) (both signals are low-pass filtered at 100~kHz). (c) Power spectrum
    of the bolometer signal with (red) and without (grey)
    feedback. Inset: same data with logarithmic vertical scale. Feedback delay $\tau=0.2~$ms.}
  
\label{fig:3}
\end{figure}

\begin{figure*}[htbp]\center
  \includegraphics[width=1\linewidth]{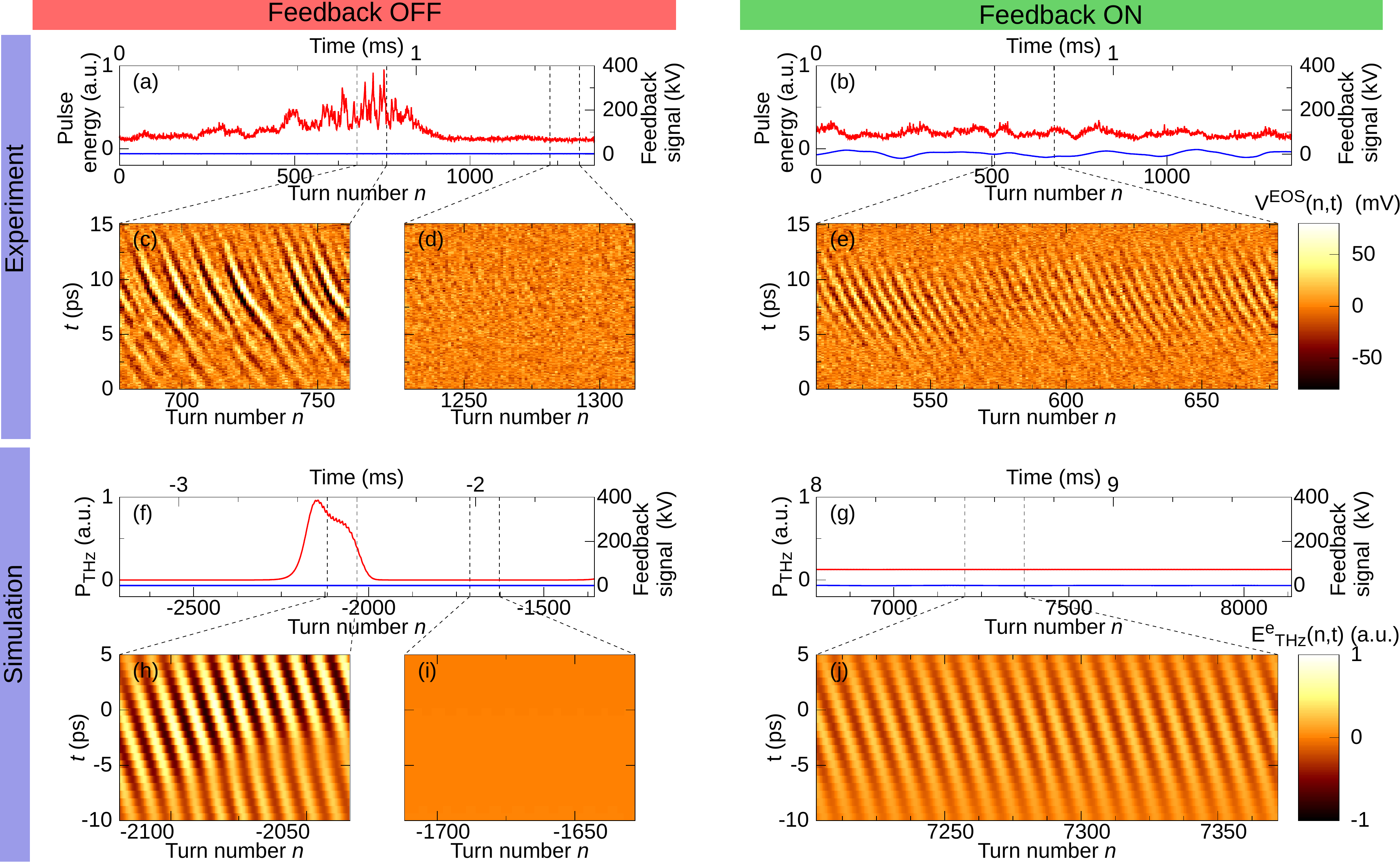}
  \caption{{\bf Evolution of the  coherent terahertz pulse shapes with and without control} (right and left respectively).
    During the experiment (top), these terahertz pulse shapes are recorded using a high   repetition rate single-shot electro-optic sampling setup based on
  the photonic time-stretch technique (see Methods). The colorscale images (c-e) represent the emitted coherent terahertz pulse shape $V^{EOS}(n,t)$   (electric field) versus time $t$ and number of turns $n$ in the ring. The red and blue curve represent the terahertz pulse energy versus the number of turns  (integration over the time of the square of the EOS signal), and the control signal (modification of the power RF amplitude). Results from simulations (bottom) are from the same data set than in Fig.~2 (see Methods for details).}
\label{fig:4}
\end{figure*}
We have tested experimentally this feedback scheme at the SOLEIL synchrotron radiation facility (Figure~\ref{fig:3}). During the
experiments, our storage ring was operated as for normal user operation, at an energy of 2.75~GeV (see Methods for parameters). We
stored a single high charge electron bunch ($I=9.15$~mA) in order to exceed the microbunching instability threshold (at about 8.7~mA), and obtain
strong bursts of coherent THz emission. In each feedback control experiment, we started by testing the control efficiency versus
parameters $G$ and $\tau$ in ranges that are suggested by numerical results. Automatic scans provided optimal values of $G$ and $\tau$ in
typically 15~minutes. Typical feedback control results are represented in Figure~\ref{fig:3}. Figures~\ref{fig:3}(a) and \ref{fig:3}(b) display the
transient obtained when the feedback control is suddenly switched ON and OFF respectively. The Fourier spectra of the controlled and
uncontrolled signals (Fig.~\ref{fig:3}c) show that the main spectral components of fluctuations are suppressed until more than 40~dB (for the feedback efficiency versus frequency, see the figure in the supplementary material).

Quantitative analysis of the remaining control fluctuations $\Delta V(t)$ also provide important information on the required RF amplitude
that is necessary to maintain   succesfull control of the unstable regular state (the remaining fluctuation visible in
Fig.~\ref{fig:3}). Quantitatively, the fluctuations of the RF signal amplitude is about of 8~kV RMS (over the 10~kHz bandwidth of the power RF system).
This value represents a variation of only  0.3\% of the  total RF signal slope seen by the electrons. 

In a third step we have performed an experiment to observe directly the spatio-temporal evolution of the electron-bunch
structures in the controlled and uncontrolled cases. For this purpose, we have used a single-shot electro-optic sampling system based
on photonic time-stretch, which is described elsewhere~\cite{szwaj2016high}. This system enables to record in a single-shot mode the electric field of
the THz pulses (envelope and carrier), at each storage ring turn. Figure~\ref{fig:4} shows the turn-by-turn evolution
of the THz pulses with and without feedback.  As in the simulations, without feedback there is an alternance of quiet periods (without
micro-structures) and irregular terahertz radiation. With feedback, the micro-structures appear almost always present and form a
regular pattern, i.e. they appear periodically at the tail of the bunch and drift at constant velocity toward the head (due to the rotation of the bunch in the phase-space).
Hence, the feedback not only removes the bursting behavior, but also make the micro-structures appear and propagate regularly in the bunch.

In conclusion, we reveal the existence of a regular, but unstable, dynamical state of  high-charge electron-bunches  circulating in a storage ring, and we demonstrate the possibility of stabilizing this state thanks to a low power feedback loop. As a consequence of this method (inspired by chaos of control theory), a modification of less than 0.3\% in the system parameters permits to decrease by more than 40 dB the THz fluctuations at the synchrotron SOLEIL (while keeping the coherent emission).
A foreseen application concerns the production of a stable coherent THz synchrotron radiation (CSR) in more synchrotron radiation facilities in the world, and/or during a large fraction of the year (for instance at SOLEIL).  Moreover, it is important to note that the simple feedback scheme used here was destined to a proof-of-principle experiment, and that more efficient feedback control schemes probably remain to be discovered.
Current open questions concern the possibility to achieve control at very high current and/or when the instability is characterized by many unstable eigenvalues (i.e., where Pyragas control is observed to fail). These next steps are expected to require cross-disciplinary works between the fields of accelerator physics, and nonlinear dynamics of spatially extended systems.

\section*{Methods}
\subsection*{Storage ring}
The SOLEIL storage ring is operated in single-bunch mode, at 2.75~GeV,
in the so-called "nominal-alpha mode", i.e., with a momentum
compaction factor $4.16\times 10^{-4}$.  Four superconducting
cavities are fed by power RF sources at 352.2~MHz (i.e., the
416$^{th}$ harmonic of the revolution frequency). Three cavities are
used for normal acceleration at each turn, with an accelerating
voltage amplitude of 800~kV. The fourth cavity is driven by the feedback
system (see below).

Furthermore, the storage ring is operated in the so-called {\it top-up} mode,
consisting in periodically injecting electrons for compensating the
natural losses in the ring. This mode
was a crucial point as it allowed systematic studies (as finding the
optimal parameters), without change of the storage ring current.

 \subsection*{Numerical simulations}
Numerical simulations are performed by using the one-dimensional
Vlasov-Fokker-Planck model of storage rings~\cite{Warnock:2006qa}, which is known
to reproduce the microbunching instability~\cite{schonfeldt2017parallelized,roussel2015.EOS}.
The longitudinal electron bunch dynamics with feedback is modeled by the equation:
\begin{eqnarray} \textstyle
  \frac{\partial f(q,p,\theta)}{\partial \theta} &-& \textstyle  p \frac{\partial f}{\partial q} +  \frac{\partial f}{\partial p} \left[ q (1 + \Delta V(\theta))   - I_c E^e_{\mathrm{THz}} \right]\\
  &=& 2\epsilon \frac{\partial}{\partial p}\left( p f + \frac{\partial f }{\partial p}\right)
 \end{eqnarray}
where $q$ is the normalized longitudinal position ($q=\frac{z}{\sigma_z}$
with $z$ the longitudinal coordinate and $\sigma_z$ the equilibrium RMS
bunch length without collective effect), $p$ the normalized
longitudinal energy ($p=\frac{E-E_0}{\sigma_E}$ with $E$ the electron
energy in eV, $E_0$ the storage ring nominal energy and $\sigma_E$ the
energy spread at equilibrium, without collective effect).  $\theta$ is
the normalized time $\theta=\frac{t}{2\pi f_s}$, with $t$ the time and
$f_s$ the synchrotron frequency.  $\epsilon = 1/(2\pi f_s \tau_s )$,
where $\tau_s$ is the synchrotron damping time.  The interactions
between electrons are modeled in the term $E^{B}_{\mathrm{THz}}$, where the standard
parallel plate model (with a radius $R_c$ and a distance between the plates of $2\times h$) is used (see~\cite{evain2017direct} for more details). $I_c$ is the
normalized bunch current, $I_c=I \frac{e 2\pi R_c}{2\pi f_s \sigma_E  T_0}$, with $I$ is the average beam current, $R_c$ the dipole radius of curvature, and $T_0$ the revolution period, $e$ the electron
charge. All parameters are in MKS units.
As in the experiment, the feedback is:

\begin{eqnarray}
  && \frac{dX(\theta)}{d\theta}=\frac{1}{\tau_{LP}}\left[P_{\mathrm{THz}}(\theta)-X(\theta)\right]\\
  && \Delta V(\theta)= G \left[ X(\theta) - X(\theta-\tau_{\theta}) \right], 
\end{eqnarray}
With $\tau_{LP}$ the low-pass filter value, $G$ the feedback gain, $\tau_{\theta}$ the feedback delay and $P_{\mathrm{THz}}$ the emitted THz power ($P_{\mathrm{THz}}(\theta)=\int_{k0}^{\infty} |\tilde{\rho}(q, \theta)|^2 dq$ with $\tilde{\rho}(k)$ the Fourier transform of the charge density $\rho(q)=\int_{-\infty}^{+\infty} f(q,p) dp$ and $k_0$ represent the bolometer detection cutoff).
As in the experiment, we put a limitation $\Delta V_{max}$ on the maximum value of $|\Delta V|$.
The Vlasov-Fokker Planck equation is integrated using the Warnock scheme~\cite{Warnock:2006qa}, that we have implemented as a parallel code (in MPI). \\

The parameters used in the article (synchrotron SOLEIL and feedback parameters)
are : $E_0=2.75~$GeV, $\sigma_z=4.59~$mm, $\sigma_E=1.017\times 10^{-3} E_0$, $f_s=4.64~$kHz, $\tau_s=3.27~$ms, $R_c=5.36~$m, $T_0=1.181~\mu$s, $h=1.25~$cm, $I=9.15~$mA,  $k_0=2.5$, $G=-0.3$, $\tau_{\theta}=5.83$ (0.2~ms in non normalized units), $1/\tau_{LP}=0.105$ (3~kHz in non normalized units), and $\Delta V_{max}= 180$~kV. 

\subsection*{Feedback control system: detection and low power electronics}
The THz signal is monitored by an InSb hot electron
  bolometer (Infrared Laboratories), with $1$~$\mu$s response
time, and AC output coupling. The bolometer detects the Terahertz
Coherent Synchrotron Radiation emitted at the AILES beamline of
synchrotron SOLEIL. The bolometer signal is then digitized and
processed using a low-cost Field-Programmable Gate Array (FPGA) board
(Red Pitaya STEMlab 125-14 board, based on the Xilinx Zynq 7010 SOC-FPGA). The
digitization is performed by one of the two ADCs of the FPGA board at
125~MS/s, with a 50~MHz bandwidth. The acquired signal is first
digitally low-pass filtered using a first order filter (with a cutoff
frequency $f_c=3$~kHz), and resampled at 1~MS/s. The FPGA uses this
filtered signal $X(t)$ to compute the feedback signal $\Delta
V=G\left[X(t) - X(t-\tau)\right]$. This digital signal is then
converted to an analog signal, using one of the two DACs of the
FPGA board.

In addition, the bolometer and control signals are also monitored
using a 1~GHz oscilloscope (Lecroy WR104MXI). The bolometer,
oscilloscope, FPGA (and its control computer) are placed in the AILES
beamline area, and the analog control signal provided by the FPGA is
transported to the low-level RF (LLRF) system of SOLEIL located at few
tens of meters from the FPGA, using a coaxial cable. This signal is used to
modulate the amplitude of one of the RF accelerating cavities.

\subsection*{Feedback control system: high power part}
At SOLEIL, four accelerating cavities are usually fed with 352.2~MHz
power RF. For this feedback control experiment, one of the four
cavities has been operated in the so-called "zero-crossing" mode,
i.e., the phase of the electric field is adjusted so that the field is
zero for the synchronous particle at each electron bunch passage as displayed in
Fig.~\ref{fig:2}. In absence of feedback, the cavity voltage amplitude
is 700~kV. The control signal delivered by the FPGA is limited to $\pm
1$~V, which corresponds to an RF amplitude modification of $\pm
180~$kV. The bandwidth of this amplitude modulation has been measured
to be 10~KHz.

\subsection*{Feedback control system: protection against accidental beam loss}
While searching for optimum feedback parameters $G$ and $\tau$, some
parameter sets have destabilizing effect, leading in some cases to
beam loss. In order to avoid this issue, we also added an interlock
system, that disables the feedback loop when the electron bunch
transverse position (monitored by a beam position monitor) departs the
nominal orbit by a threshold value. This value has been adjusted
empirically (by trial-and-error) to $\pm~50~\mu$m. Operation of this
interlock was an important component, as it allowed us to freely
scan feedback parameters and find the optimal ones, without needing to
take care of possible electron beam loss issues.

\subsection*{Single-shot recording of the terahertz CSR pulses}
The data displayed in Figure~\ref{fig:4} (emitted THz pulse shapes at
each turn) have been recorded using a single-shot electro-optic
sampling system. The setup, which is detailed in
Refs.~\cite{szwaj2016high,evain2017direct}, uses a single-shot
electro-optic sampling using chirped laser pulses, combined
with photonic time-stretch~\cite{mahjoubfar2017time} for enabling high repetition rate
recording.

\section*{Data availability}
The data that support the findings of this study are available from the
corresponding author upon reasonable request.

\section*{Author contributions}
S.B. and C.E. carried out the numerical simulations. C.E., S.B. and C.S. led the experimental realization.
C.E., S.B. and J.R. developed the FPGA software.
Experiments at SOLEIL designed and performed by M.-A.T. (ring configuration and operation), F.R. (RF system configuration and seetings), M. Labat and N. Hubert (interlock and diagnostic systems), J.-B. B. and P. Roy (AILES beamline),   M.L.-P., E.R., S.B.,C.E, C.S. (Electro-Optics-Sampling detection system and feedback system).
Experimental data analysed by C.S. and C.E.
All the authors participate to the redaction.

\section*{Acknowledgments}
This work has been partially supported by the LABEX CEMPI (ANR-11-LABX-0007) and the Equipex Flux (ANR-11-EQPX-0017), as well as
by the Ministry of Higher Education and Research, Hauts de  France council and European Regional Development Fund (ERDF) through the
Contrat de Projets Etat-Region (CPER Photonics for Society P4S). The project used HPC resources from GENCI TGCC/IDRIS (i2016057057,A0040507057).

\bibliographystyle{naturemag}

\begin{thebibliography}{10}
\expandafter\ifx\csname url\endcsname\relax
  \def\url#1{\texttt{#1}}\fi
\expandafter\ifx\csname urlprefix\endcsname\relax\def\urlprefix{URL }\fi
\providecommand{\bibinfo}[2]{#2}
\providecommand{\eprint}[2][]{\url{#2}}

\bibitem{charru2013sand}
\bibinfo{author}{Charru, F.}, \bibinfo{author}{Andreotti, B.} \&
  \bibinfo{author}{Claudin, P.}
\newblock \bibinfo{title}{Sand ripples and dunes}.
\newblock \emph{\bibinfo{journal}{Annual Review of Fluid Mechanics}}
  \textbf{\bibinfo{volume}{45}}, \bibinfo{pages}{469--493}
  (\bibinfo{year}{2013}).

\bibitem{hopkin2004sea}
\bibinfo{author}{Hopkin, M.}
\newblock \bibinfo{title}{Sea snapshots will map frequency of freak waves}.
\newblock \emph{\bibinfo{journal}{Nature}}  (\bibinfo{year}{2004}).

\bibitem{solli2007optical}
\bibinfo{author}{Solli, D.}, \bibinfo{author}{Ropers, C.},
  \bibinfo{author}{Koonath, P.} \& \bibinfo{author}{Jalali, B.}
\newblock \bibinfo{title}{Optical rogue waves}.
\newblock \emph{\bibinfo{journal}{Nature}} \textbf{\bibinfo{volume}{450}},
  \bibinfo{pages}{1054} (\bibinfo{year}{2007}).

\bibitem{helbing2001traffic}
\bibinfo{author}{Helbing, D.}
\newblock \bibinfo{title}{Traffic and related self-driven many-particle
  systems}.
\newblock \emph{\bibinfo{journal}{Reviews of modern physics}}
  \textbf{\bibinfo{volume}{73}}, \bibinfo{pages}{1067} (\bibinfo{year}{2001}).

\bibitem{abo2002.PhysRevLett.88.254801}
\bibinfo{author}{Abo-Bakr, M.}, \bibinfo{author}{Feikes, J.},
  \bibinfo{author}{Holldack, K.}, \bibinfo{author}{W\"ustefeld, G.} \&
  \bibinfo{author}{H\"ubers, H.-W.}
\newblock \bibinfo{title}{Steady-state far-infrared coherent synchrotron
  radiation detected at {BESSY {II}}}.
\newblock \emph{\bibinfo{journal}{Phys. Rev. Lett.}}
  \textbf{\bibinfo{volume}{88}}, \bibinfo{pages}{254801}
  (\bibinfo{year}{2002}).

\bibitem{byrd2002.PhysRevLett.89.224801}
\bibinfo{author}{Byrd, J.~M.} \emph{et~al.}
\newblock \bibinfo{title}{Observation of broadband self-amplified spontaneous
  coherent terahertz synchrotron radiation in a storage ring}.
\newblock \emph{\bibinfo{journal}{Phys. Rev. Lett.}}
  \textbf{\bibinfo{volume}{89}}, \bibinfo{pages}{224801}
  (\bibinfo{year}{2002}).

\bibitem{venturini2002.PhysRevLett.89.224802}
\bibinfo{author}{Venturini, M.} \& \bibinfo{author}{Warnock, R.}
\newblock \bibinfo{title}{Bursts of coherent synchrotron radiation in electron
  storage rings: A dynamical model}.
\newblock \emph{\bibinfo{journal}{Phys. Rev. Lett.}}
  \textbf{\bibinfo{volume}{89}}, \bibinfo{pages}{224802}
  (\bibinfo{year}{2002}).

\bibitem{uvsor_MBI_ybco}
\bibinfo{author}{Roussel, E.} \emph{et~al.}
\newblock \bibinfo{title}{Microbunching instability in relativistic electron
  bunches: Direct observations of the microstructures using ultrafast {YBCO}
  detectors}.
\newblock \emph{\bibinfo{journal}{Phys. Rev. Lett.}}
  \textbf{\bibinfo{volume}{113}}, \bibinfo{pages}{094801}
  (\bibinfo{year}{2014}).

\bibitem{roussel2015.EOS}
\bibinfo{author}{Roussel, E.} \emph{et~al.}
\newblock \bibinfo{title}{Observing microscopic structures of a relativistic
  object using a time-stretch strategy}.
\newblock \emph{\bibinfo{journal}{Scientific Reports}}
  \textbf{\bibinfo{volume}{5}} (\bibinfo{year}{2015}).

\bibitem{brosi2016fast}
\bibinfo{author}{Brosi, M.} \emph{et~al.}
\newblock \bibinfo{title}{Fast mapping of terahertz bursting thresholds and
  characteristics at synchrotron light sources}.
\newblock \emph{\bibinfo{journal}{Physical Review Accelerators and Beams}}
  \textbf{\bibinfo{volume}{19}}, \bibinfo{pages}{110701}
  (\bibinfo{year}{2016}).

\bibitem{PhysRevAccelBeams.19.020704}
\bibinfo{author}{Billinghurst, B.~E.} \emph{et~al.}
\newblock \bibinfo{title}{Longitudinal bunch dynamics study with coherent
  synchrotron radiation}.
\newblock \emph{\bibinfo{journal}{Phys. Rev. Accel. Beams}}
  \textbf{\bibinfo{volume}{19}}, \bibinfo{pages}{020704}
  (\bibinfo{year}{2016}).

\bibitem{ott1990controlling}
\bibinfo{author}{Ott, E.}, \bibinfo{author}{Grebogi, C.} \&
  \bibinfo{author}{Yorke, J.~A.}
\newblock \bibinfo{title}{Controlling chaos}.
\newblock \emph{\bibinfo{journal}{Physical review letters}}
  \textbf{\bibinfo{volume}{64}}, \bibinfo{pages}{1196} (\bibinfo{year}{1990}).

\bibitem{shinbrot1993using}
\bibinfo{author}{Shinbrot, T.}, \bibinfo{author}{Grebogi, C.},
  \bibinfo{author}{Yorke, J.~A.} \& \bibinfo{author}{Ott, E.}
\newblock \bibinfo{title}{Using small perturbations to control chaos}.
\newblock \emph{\bibinfo{journal}{nature}} \textbf{\bibinfo{volume}{363}},
  \bibinfo{pages}{411} (\bibinfo{year}{1993}).

\bibitem{pyragas1992continuous}
\bibinfo{author}{Pyragas, K.}
\newblock \bibinfo{title}{Continuous control of chaos by self-controlling
  feedback}.
\newblock \emph{\bibinfo{journal}{Physics letters A}}
  \textbf{\bibinfo{volume}{170}}, \bibinfo{pages}{421--428}
  (\bibinfo{year}{1992}).

\bibitem{stupakov2002.PhysRevSTAB.5.054402}
\bibinfo{author}{Stupakov, G.} \& \bibinfo{author}{Heifets, S.}
\newblock \bibinfo{title}{Beam instability and microbunching due to coherent
  synchrotron radiation}.
\newblock \emph{\bibinfo{journal}{Phys. Rev. ST Accel. Beams}}
  \textbf{\bibinfo{volume}{5}}, \bibinfo{pages}{054402} (\bibinfo{year}{2002}).

\bibitem{sannibale2004.PhysRevLett.93.094801}
\bibinfo{author}{Sannibale, F.} \emph{et~al.}
\newblock \bibinfo{title}{A model describing stable coherent synchrotron
  radiation in storage rings}.
\newblock \emph{\bibinfo{journal}{Phys. Rev. Lett.}}
  \textbf{\bibinfo{volume}{93}}, \bibinfo{pages}{094801}
  (\bibinfo{year}{2004}).

\bibitem{Warnock:2006qa}
\bibinfo{author}{Warnock, R.~L.}
\newblock \bibinfo{title}{{Study of bunch instabilities by the nonlinear
  Vlasov-Fokker-Planck equation}}.
\newblock \emph{\bibinfo{journal}{Nucl. Instrum. Meth. Phys. Res., Sect. A}}
  \textbf{\bibinfo{volume}{561}}, \bibinfo{pages}{186--194}
  (\bibinfo{year}{2006}).

\bibitem{shimada2009transverse}
\bibinfo{author}{Shimada, M.} \emph{et~al.}
\newblock \bibinfo{title}{Transverse-longitudinal coupling effect in laser
  bunch slicing}.
\newblock \emph{\bibinfo{journal}{Physical review letters}}
  \textbf{\bibinfo{volume}{103}}, \bibinfo{pages}{144802}
  (\bibinfo{year}{2009}).

\bibitem{feikes2011metrology}
\bibinfo{author}{Feikes, J.} \emph{et~al.}
\newblock \bibinfo{title}{Metrology light source: The first electron storage
  ring optimized for generating coherent thz radiation}.
\newblock \emph{\bibinfo{journal}{Physical Review Special Topics-Accelerators
  and Beams}} \textbf{\bibinfo{volume}{14}}, \bibinfo{pages}{030705}
  (\bibinfo{year}{2011}).

\bibitem{martin2011experience}
\bibinfo{author}{Martin, I.}, \bibinfo{author}{Rehm, G.},
  \bibinfo{author}{Thomas, C.} \& \bibinfo{author}{Bartolini, R.}
\newblock \bibinfo{title}{Experience with low-alpha lattices at the diamond
  light source}.
\newblock \emph{\bibinfo{journal}{Physical Review Special Topics-Accelerators
  and Beams}} \textbf{\bibinfo{volume}{14}}, \bibinfo{pages}{040705}
  (\bibinfo{year}{2011}).

\bibitem{roy.RSI.84.033102.2013}
\bibinfo{author}{Barros, J.} \emph{et~al.}
\newblock \bibinfo{title}{Coherent synchrotron radiation for broadband
  terahertz spectroscopy}.
\newblock \emph{\bibinfo{journal}{Review of Scientific Instruments}}
  \textbf{\bibinfo{volume}{84}}, \bibinfo{pages}{033102}
  (\bibinfo{year}{2013}).

\bibitem{tammaro2015high}
\bibinfo{author}{Tammaro, S.} \emph{et~al.}
\newblock \bibinfo{title}{High density terahertz frequency comb produced by
  coherent synchrotron radiation}.
\newblock \emph{\bibinfo{journal}{Nature communications}}
  \textbf{\bibinfo{volume}{6}} (\bibinfo{year}{2015}).

\bibitem{steinmann2016frequency}
\bibinfo{author}{Steinmann, J.~L.} \emph{et~al.}
\newblock \bibinfo{title}{Frequency-comb spectrum of periodic-patterned
  signals}.
\newblock \emph{\bibinfo{journal}{Physical review letters}}
  \textbf{\bibinfo{volume}{117}}, \bibinfo{pages}{174802}
  (\bibinfo{year}{2016}).

\bibitem{boccaletti2000control}
\bibinfo{author}{Boccaletti, S.}, \bibinfo{author}{Grebogi, C.},
  \bibinfo{author}{Lai, Y.-C.}, \bibinfo{author}{Mancini, H.} \&
  \bibinfo{author}{Maza, D.}
\newblock \bibinfo{title}{The control of chaos: theory and applications}.
\newblock \emph{\bibinfo{journal}{Physics reports}}
  \textbf{\bibinfo{volume}{329}}, \bibinfo{pages}{103--197}
  (\bibinfo{year}{2000}).

\bibitem{Ahlborn2004}
\bibinfo{author}{Ahlborn, A.} \& \bibinfo{author}{Parlitz, U.}
\newblock \bibinfo{title}{Stabilizing unstable steady states using multiple
  delay feedback control}.
\newblock \emph{\bibinfo{journal}{Phys. Rev. Lett.}}
  \textbf{\bibinfo{volume}{93}}, \bibinfo{pages}{264101}
  (\bibinfo{year}{2004}).

\bibitem{roy.IPT.49.139.2006}
\bibinfo{author}{Roy, P.}, \bibinfo{author}{Rouzi\`eres, M.},
  \bibinfo{author}{Qi, Z.} \& \bibinfo{author}{Chubar, O.}
\newblock \bibinfo{title}{The ailes infrared beamline on the third generation
  synchrotron radiation facility soleil}.
\newblock \emph{\bibinfo{journal}{Infrared Physics and Technology}}
  \textbf{\bibinfo{volume}{49}}, \bibinfo{pages}{139} (\bibinfo{year}{2006}).

\bibitem{evain2017direct}
\bibinfo{author}{Evain, C.} \emph{et~al.}
\newblock \bibinfo{title}{Direct observation of spatiotemporal dynamics of
  short electron bunches in storage rings}.
\newblock \emph{\bibinfo{journal}{Physical Review Letters}}
  \textbf{\bibinfo{volume}{118}}, \bibinfo{pages}{054801}
  (\bibinfo{year}{2017}).

\bibitem{szwaj2016high}
\bibinfo{author}{Szwaj, C.} \emph{et~al.}
\newblock \bibinfo{title}{High sensitivity photonic time-stretch electro-optic
  sampling of terahertz pulses}.
\newblock \emph{\bibinfo{journal}{Review of Scientific Instruments}}
  \textbf{\bibinfo{volume}{87}}, \bibinfo{pages}{103111}
  (\bibinfo{year}{2016}).

\bibitem{schonfeldt2017parallelized}
\bibinfo{author}{Sch{\"o}nfeldt, P.}, \bibinfo{author}{Brosi, M.},
  \bibinfo{author}{Schwarz, M.}, \bibinfo{author}{Steinmann, J.~L.} \&
  \bibinfo{author}{M{\"u}ller, A.-S.}
\newblock \bibinfo{title}{Parallelized vlasov-fokker-planck solver for desktop
  personal computers}.
\newblock \emph{\bibinfo{journal}{Physical Review Accelerators and Beams}}
  \textbf{\bibinfo{volume}{20}}, \bibinfo{pages}{030704}
  (\bibinfo{year}{2017}).

\bibitem{mahjoubfar2017time}
\bibinfo{author}{Mahjoubfar, A.} \emph{et~al.}
\newblock \bibinfo{title}{Time stretch and its applications}.
\newblock \emph{\bibinfo{journal}{Nature Photonics}}
  \textbf{\bibinfo{volume}{11}}, \bibinfo{pages}{341--351}
  (\bibinfo{year}{2017}).

\end{thebibliography}

\end{document}